\documentstyle[preprint,aps,epsfig,tighten]{revtex}

\begin{document}
\draft
%\twocolumn[\hsize\textwidth\columnwidth\hsize\csname@twocolumnfalse%
%\endcsname
\preprint{\parbox[t]{80mm}{\small Zagreb University preprint 
ZTF-99/08}}

\title{The quark loop calculation of the 
$\gamma \to 3 \pi$ form factor }

\author{Bojan Bistrovi\' c and Dubravko Klabu\v{c}ar}
\address{\footnotesize Department of Physics, Faculty of Science, \\
        Zagreb University, P.O.B. 162, 10001 Zagreb, Croatia}
%\date\today
\maketitle

\begin{abstract}
The presently experimentally interesting 
form factor for the anomalous process $\gamma \to \pi^+ \pi^0 \pi^-$ 
is calculated as the quark ``box"-amplitude where the intermediate
fermion loop is the one of constituent quarks with the pseudoscalar
coupling to pions. This also corresponds to the form factor, in the 
lowest order in pion interactions, of the $\sigma$-model and of the 
chiral quark model. We give the analytic expression for the form 
factor in terms of an expansion in the pion momenta up to the order 
${\cal O}(p^8)$ relative to the soft point result, and also perform 
its exact numerical evaluation. We compare our predictions with 
those of the vector meson dominance and chiral perturbation theory, 
as well as with the scarce data available so far.

%\vspace{3mm}
%\noindent {\it Keywords:} Axial anomaly; Quark loops;
%Box graph; Constituent quarks.

\end{abstract}

\vspace{3mm}
%]

%\section{Introduction}

The analysis of the Abelian axial anomaly \cite{Adler69,BellJackiw69}
shows that the $\pi^0\to\gamma\gamma$ amplitude is exactly
$T_{\pi}^{2\gamma}(m_\pi = 0) = e^2 N_c /(12\pi^2 f_\pi)$
in the chiral and soft limit of pions of vanishing mass $m_\pi$.
(Here, $e$ is the proton charge, and $N_c=3$ is the number of quark 
colors.) 
This amplitude $T_{\pi}^{2\gamma}(m_\pi = 0)$ is successfully 
reproduced also by the simple ``free" quark loop (QL) calculation
of the pseudoscalar-vector-vector (PVV) ``triangle",
provided one uses the quark-level Goldberger-Treiman (GT) relation
$g/M = 1/f_{\pi}$ to express the effective constituent quark mass 
$M$ and quark-pion coupling strength $g$ in terms of the pion decay 
constant $f_{\pi} = 92.4$ MeV. This calculation (essentially 
{\it {\` a} la} Steinberger \cite{S49}) is the same as the lowest 
(one-loop) order calculation \cite{BellJackiw69} in the 
$\sigma$-model which was constructed to realize current algebra 
explicitly. By ``free'' quarks we mean that there are no interactions
between the effective constituent quarks in the loop, while they 
{\it do} couple to external fields, presently the photons $A_\mu$ 
and the pion $\pi_a$. Our effective QL model Lagrangian is thus
\begin{equation}
{\mathcal{L}}_{\mathit{eff}} = 
\bar\Psi\left( i \partial \hskip-0.55em{\slash} 
 - e {\cal Q} A \hskip-0.5em\slash -M\right)\Psi
- i \,g\; \bar\Psi \gamma_5 \pi_a\tau_a\; \Psi + ... \; ,
\label{1Leff}
\end{equation}
where ${\cal Q}\equiv\mbox{\rm diag}(Q_u,Q_d)= \mbox{\rm diag}
(\frac{2}{3},-\frac{1}{3})$ is the quark charge 
matrix, and $\tau_a$ are the Pauli $SU(2)$-isospin matrices acting 
on the quark iso-doublets $\Psi = (u, d)^T$. The ellipsis in 
${\mathcal{L}}_{\mathit{eff}}$ 
serve to remind us that Eq. (\ref{1Leff}) also represents the 
lowest order terms from the $\sigma$-model Lagrangian which are 
pertinent for calculating photon-pion processes. The same holds 
for all chiral quark models ($\chi$QM) -- considered in, 
{\it e.g.}, Ref. \cite{Andrianov+al98} -- which
contain quark-meson coupling $M{\bar \Psi}(UP_L+U^\dagger P_R)\Psi$
with $P_{L,R} = (1\pm \gamma_5)/2$. Namely, expanding 
$U^{(\dagger)}=\exp[(-)i \pi_a\tau_a/f_\pi]$ to the lowest order in 
$\pi_a$ and invoking the GT relation, again returns the QL model 
Lagrangian (\ref{1Leff}). 

This simple QL model (and hence also the lowest order $\chi$QM and
the $\sigma$-model) provides an analytic expression ({\it e.g.}, see 
Ref. \cite{Ametller+al83}) for the amplitude 
$T_{\pi}^{2\gamma}(m_\pi)$ 
also for $m_\pi > 0$ (but restricted to $m_\pi < 2 M$, which anyway 
must hold for the light, pseudo-Goldstone pion), namely
\begin{equation}
T_{\pi}^{2\gamma}(m_\pi) = \frac{e^2 N_c}{12\pi^2f_\pi}
\left[ \frac{\arcsin(m_\pi/2M)}{(m_\pi/2M)}\right]^2 =
\frac{e^2 N_c}{12\pi^2f_\pi} \left[ 1 + \frac{m_\pi^2}{12M^2} +
... \right] \, .
\label{freeLoopAmp}
\end{equation}

Adler {\it et al.}, Terentev, and Aviv and Zee 
\cite{Ad+al71Te72Av+Z72} proved that the amplitude 
for the anomalous processes of the type $\gamma \to 3\pi$ 
is related to $T_{\pi}^{2\gamma}(0)$ and is given by
\begin{equation}
F_\gamma^{3\pi}(0,0,0) \, = 
\, \frac{1}{e f_\pi^2} \, T_{\pi}^{2\gamma}(0) \, =
\, \frac{e N_c}{12 \pi^2 f_\pi^3} \, .
\label{g3piAnomAmp}
\end{equation}
The arguments of the anomalous amplitude (\ref{g3piAnomAmp}),
namely the momenta $\{ p_1,p_2,p_3 \}$ of the three pions 
$\{\pi^+,\pi^0, \pi^-\}$,
are all set to zero, because Eq.~(\ref{g3piAnomAmp}) is also a soft 
limit and chiral limit result, giving the form factor 
$F_\gamma^{3\pi}(p_1,p_2,p_3)$ at the soft point.

In the QL model, the amplitude (\ref{g3piAnomAmp}) is
obtained by calculating the ``box" graph, Fig. 1. 
This is not surprising,
as the anomalous ``box" amplitude (\ref{g3piAnomAmp}) was already
obtained analytically and exactly by Alkofer and Roberts \cite{AR96}
in the so-called Schwinger-Dyson (SD) {\it Ansatz} approach, which is
more general. In this approach, one employs an {\it Ansatz} for a
dynamically dressed quark propagator, characterized by a
momentum-dependent quark mass function $M(k^2)$, and by the related
momentum-dependent quark-antiquark pseudoscalar pion bound state 
Bethe-Salpeter vertices $\Gamma_{\pi^a}$ instead of our 
$g\gamma_5\tau_a$ quark-pion Yukawa couplings. 
The present free QL model, with its constant 
quark-pion coupling strength $g$ and the free quark propagators 
$S(k) = i ({\gamma\cdot k+M})/({k^2-M^2})$ 
containing constant effective constituent mass $M$, can therefore be 
considered as a special case of the SD {\it Ansatz} approach, 
{\it i.e.}, the simplest possible such {\it Ansatz}.

Let us stress that for the processes of the type
$\gamma\to 3\pi$, going beyond the soft limit is much more important
than for the process $\pi^0 \to 2\gamma$ where the amplitude
$T_{\pi}^{2\gamma}$ obtained in the chiral and soft limit is an 
excellent approximation for the realistic $\pi^0\to\gamma\gamma$ 
decay. By contrast, the current TJNAF measurement of the 
$\gamma \pi^+ \to \pi^+ \pi^0$ process \cite{Miskimen+al94}, 
as well as already published \cite{Antipov+al87} and still planned 
Primakoff measurements at CERN \cite{Moinester+al99}, involve so 
large values of energy and momentum transfer, that departures from 
the soft-point result (\ref{g3piAnomAmp}) may well be significant.

Nevertheless, to the best of our knowledge, 
no extension of Eq. (\ref{g3piAnomAmp}) beyond the
chiral and soft limit has been given in the 
QL ($\chi$QM, $\sigma$-model) approach so far, 
neither through a numerical calculation, nor
as an analytic result that would be the analogy of the amplitude
(\ref{freeLoopAmp}) for the $\gamma\gamma$-decay of the massive pion.
(In the SD approach \cite{AR96}, the box graph beyond the chiral 
and soft limit was addressed, but only numerically.)
To fill this gap that seems to exist in the literature
even at the lowest non-trivial order, we provide such 
an analytic expression in the form of an expansion in 
powers of the pion momenta up to the eighth order, in 
addition to the numerical evaluation of the
$\gamma \to 3\pi$ form factor accurate to all orders.

%\section{Calculation}

To compute the $\gamma\to 3\pi$ amplitude we use Eq. (\ref{1Leff})
in terms of the physical fields $\pi^\pm=(\pi_1\mp i \pi_2)/\sqrt{2}$
and $\pi^0\equiv\pi_3$, so that $\pi_a \tau_a = \sqrt{2}(\pi^+\tau_+
+ \pi^- \tau_-) + \pi^0 \tau_3$ where 
$\tau_{\pm}=(\tau_1 \pm i\tau_2$)/2.
There are six different contributing graphs, obtained from Fig. 1 by
the permutations of the vertices of the three different pions. 
The momenta flowing through the four sections of the quark loop
are conveniently given by various combinations of the symbols
$\alpha, \beta, \gamma = +, 0, -$ in $k_{\alpha\beta\gamma} \equiv 
k + \alpha p_{1} + \beta p_{2} + \gamma p_{3}$.
Then, the anomalous vertex $V_\mu$ coupling ${\pi^+}{\pi^0}{\pi^-}$ 
to $\gamma$, ${\it viz.}$ the scalar form factor  
$F^{3\pi}_\gamma(p_{1},p_{2},p_{3})$ associated to it, is in the 
present approach calculated through the six VPPP box graphs as 
\begin{eqnarray}
V_\mu &=& i \epsilon_{\mu\nu\rho\sigma}\:
p_1^\nu p_2^\rho p_3^\sigma \:
F^{3\pi}_\gamma(p_1,p_1,p_3)\nonumber \\
&=&
i\epsilon_{\mu\nu\rho\sigma}\: p_1^\nu p_2^\rho p_3^\sigma
\:
f^{3\pi}_\gamma(p_1,p_2,p_3)
+ (\mbox{permutations of} \,\,\,\,\, \pi^+, \pi^0, \pi^-) \; ,
\label{g3piVertex}
\end{eqnarray}
where the contribution of the first diagram (Fig. 1), 
$ i\epsilon_{\mu\nu\rho\sigma}\: p_1^\nu p_2^\rho p_3^\sigma
 f^{3\pi}_\gamma(p_1,p_2,p_3)$, is given by
\begin{eqnarray}
- \int\frac{\,d^4 k}{(2\pi)^4}\, {\mathrm{Tr}} \left\{e{\mathcal{Q}} 
\gamma_\mu \; S(k) \; i\sqrt{2}\, g \gamma_5 \tau_{+} \; 
S(k_{\scriptscriptstyle -00}) \;
ig\gamma_5 \tau_3 \; S(k_{\scriptscriptstyle --0}) \;
i\sqrt{2}\,g\gamma_5 \tau_{-} \; S(k_{\scriptscriptstyle ---})
\right\} \, .
\end{eqnarray}
The color and isospin traces are $N_c$ and 
${\mathrm{Tr}}_{I}\left({\cal Q} \tau_+ \tau_3 \tau_-\right) = -2/3$,
respectively. The Dirac trace is 
$4iM \epsilon_{\mu\nu\rho\sigma}\: p_1^\nu p_2^\rho p_3^\sigma \, $,
leading to the partial amplitude 
$f^{3\pi}_\gamma(p_1,p_2,p_3) = 
(+2/3)(e N_c g^3 M/2 \pi^2) I(p_1,p_2,p_3)$, where
\begin{equation}
I(p_1,p_2,p_3) \equiv -\frac{i}{\pi^2}  \int\frac{\,d^4 k}{(k^2-M^2)
(k_{\scriptscriptstyle -00}^2-M^2)(k_{\scriptscriptstyle --0}^2-M^2) (k_{\scriptscriptstyle ---}^2-M^2)} \, .
\label{eqInt}
\end{equation}     
After combining the four propagator denominators by the Feynman 
trick, and shifting $k^\mu$ by $ p_1^\mu (1-x_3) 
+  p_2^\mu (1-x_2)  +  p_3^\mu (1- x_1) $,
the integral over the loop momentum $k$ becomes
\begin{eqnarray}  
I(p_1,p_2,p_3) &=&  -\frac{i}{\pi^2} 3!
\int _{0}^{1} \,d {x_1} \int _{0}^{x_1} 
\,d {x_2} \int _{0}^{x_2}\,d {x_3}
\int\frac{\,d^4 k}{\left[ {k^2} - {R(p_1,p_2,p_3)^2} \right]^4} 
\nonumber \\
& = & \int _{0}^{1}\int _{0}^{x_1} \int _{0}^{x_2}
\frac{\,d {x_3}\,d {x_2} \,d {x_1}}{R(p_1,p_2,p_3)^4}   
\qquad\quad {\rm (for}\,\, R^2 > 0 {\rm )} \, ,
\end{eqnarray}
where
\begin{eqnarray}
R(p_1,p_2,p_3)^2&\equiv&M^2 - p_3^2 x_1(1 - x_1)  
                       - p_2^2 x_2(1 - x_2) - p_1^2 x_3(1 - x_3)
    \nonumber \\ & & 
- 2 p_2\cdot p_3 x_2(1 -  x_1)
- 2 p_1\cdot p_2 x_3(1 - x_2 ) 
- 2 p_1\cdot p_3 x_3(1 - x_1 ) \, .
\label{defR}
\end{eqnarray}

Adding up the contributions 
from the remaining 5 diagrams yields the total amplitude 
\begin{eqnarray}
F^{3\pi}_\gamma(p_1,p_2,p_3) &=& 
\left(\frac{ e N_c }{2 \pi^2}g^3 M\right)
\left(+\frac{2}{3} \left\{
I(p_1,p_2,p_3) + I(p_1,p_3,p_2) + I(p_2,p_1,p_3) \right\}
\right. \nonumber \\ && \left.
 -\frac{1}{3} \left\{
I(p_3,p_1,p_2) + I(p_3,p_2,p_1) + I(p_2,p_3,p_1) \right\}  \right) \, .
\label{sumF}
\end{eqnarray}
Since the integrals such as $I(p_1,p_2,p_3)$ are 
symmetric in the interchange of the first and the third argument, 
the two curly brackets in Eq. (\ref{sumF}) are equal to each other. 
Therefore, the sum of the three $u$-quark ``box" loop 
diagrams, as well as the sum of the three $d$-quark ones, give 
contributions to the amplitude (\ref{sumF}) which are separately
symmetric under $p_1 \leftrightarrow p_2 \leftrightarrow p_3$.
For that reason, a calculation that would employ the integer charge
iso-doublet nucleons in the loops instead of the ones with quarks, 
would -- up to the fermion mass values, $M$ {\it vs.} $M_{nucleon}$ 
-- give the same amplitude in spite of the three neutron graphs 
dropping out because of the vanishing neutron charge. Also, the 
interplay of the charge-isospin factors compensates the quark color 
factor $N_c=3$  as in the 
$\pi^0\to \gamma \gamma$ amplitude $T^{2\gamma}_\pi$, although 
the quark charges enter in $F^{3\pi}_{\gamma}$ linearly, and not
quadratically. 

The integrals appearing in Eq. (\ref{sumF}), exemplified above by 
the one defined by Eqs. (\ref{eqInt})-(\ref{defR}), are calculated 
in two ways. 
Firstly, we calculate them numerically, by the Gauss-Kronrod method.
Secondly, we obtain the analytic expressions for $F^{3\pi}_\gamma$ 
[Eqs. (\ref{eqpi}),(\ref{eqant}), (\ref{eqmisk}) and 
(\ref{eqmisk2}) below]
by expanding the integrands [exemplified by the one in the integral
(\ref{eqInt})-(\ref{defR})] in the series of the scalar products of 
the pion momenta, $p_i\cdot p_j$ ($i,j = 1,2,3$).
In this case, the analytic integration over $x_1, x_2$ and $x_3$ 
finally yields [to order ${\mathcal{O}}(p^4)$]
\begin{eqnarray}\label{eqpi}
F^{3\pi}_\gamma(p_1,p_2,p_3)&=&
\left(\frac{ e N_c }{12 \pi^2}\frac{g^3}{ M^3} \right)
\left(
1+
  \frac{1}{3\,M^2} \left\{ p_1\cdot p_2+ p_1\cdot p_3
        + p_2\cdot p_3+ p_1^2+ p_2^2+ p_3^2 \right\}
\right. \nonumber \\ & & \left.
	 + \frac{1}{10\,M^4}\left\{ p_1^4 + p_2^4 + p_3^4 \right\}
	 + \frac{1}{6\,M^4}\left\{ p_1^2\,p_2^2 
         + p_1^2\,p_3^2 + p_2^2\,p_3^2 \right\}
\right. \nonumber \\ & & \left.
	+ \frac{1}{5\,M^4}\left\{ p_1\cdot p_2\,(p_1^2+ p_2^2)
	+ p_2\cdot p_3\,( p_2^2+p_3^2) 
        + p_1\cdot p_3\, (p_1^2+ p_3^2) \right\}
\right. \nonumber \\ & & \left.
	+ \frac{1}{6\,M^4}\left\{ p_2\cdot p_3\,p_1^2 
        + p_1\cdot p_3\,p_2^2
	+ p_1\cdot p_2\,p_3^2 \right\}
\right. \nonumber \\ & & \left.
	+ \frac{1}{5\,M^4}\left\{ p_1\cdot p_2\, p_1\cdot p_3
	+ p_1\cdot p_2\, p_2\cdot p_3 
        + p_1\cdot p_3\, p_2\cdot p_3 \right\}
\right. \nonumber \\ & & \left.
	+ \frac{2}{15\,M^4} \left\{ \left( p_1\cdot p_2 \right)^2
	+ \left( p_1\cdot p_3 \right)^2 
        + \left( p_2\cdot p_3 \right)^2 \right\}
+ {\mathcal{O}}(p^6) \right) \, .
\end{eqnarray}
After using the GT relation, Eq. (\ref{eqpi}) 
returns at the soft point ($p_i = 0$) the axial anomaly result 
(\ref{g3piAnomAmp}). We introduce the form factor normalized 
to the anomaly amplitude (\ref{g3piAnomAmp}), 
${\tilde{F}}^{3\pi}_\gamma(p_1,p_2,p_3) = 
F^{3\pi}_\gamma(p_1,p_2,p_3)/F^{3\pi}_\gamma(0,0,0)$.
It is also convenient to re-express the scalar products 
$p_i\cdot p_j$ through the Mandelstam variables as defined 
by Ref. \cite{Miskimen+al94}:
$s=(p_1+p_2)^2$, $t'=(p_2+p_3)^2$, $u=(p_1+p_3)^2$, while $t=p_3^2$ 
serves as the measure of virtuality of the third pion which may be 
off shell. The photon momentum is $q = p_1 + p_2 + p_3$.

We obtained the expansion for $F^{3\pi}_\gamma(p_1,p_2,p_3)$ 
to the order ${\mathcal{O}}(p^8)$ [relative to the anomaly 
result (\ref{g3piAnomAmp})], but give Eq. (\ref{eqpi}) only 
to the order ${\mathcal{O}}(p^4)$ for brevity
as the ${\mathcal{O}}(p^8)$-expansion for general $p_i$ is very
lengthy. It will be given elsewhere. However, below we do give 
$F^{3\pi}_\gamma$ to the order ${\mathcal{O}}(p^8)$ for the 
simpler special cases which are important for comparing our 
predictions with the experiments 
\cite{Antipov+al87,Miskimen+al94,Moinester+al99}. Namely, one 
can take the photon to be on shell in all three pertinent 
experiments \cite{Antipov+al87,Miskimen+al94,Moinester+al99},
in which at least two pion momenta, those of $\pi^+$ and $\pi^0$,
are also on the mass-shell. 
We thus set $q^2=0$ and $p_{1}^2=p_{2}^2 = m_\pi^2$, whereby
\begin{equation}\label{oscond}
s+t'+u \equiv p_1^2 + p_2^2 + p_3^2 + q^2 = 
2\, m_\pi^2 + t  \, .
\end{equation}

In the Primakoff measurements \cite{Antipov+al87,Moinester+al99}, 
including the one \cite{Antipov+al87} providing the only 
existing data point so far, the third pion is also on shell.
Hence, $t=m_\pi^2$, in which case we predict 
\begin{eqnarray}\label{eqant}
\widetilde{F}^{3\pi}_\gamma(s,t')&=&\left(1 + 
       \frac{m_\pi^2}{2\,M^2} + \frac{m_\pi^4}{4\,M^4} + 
\frac{169\, m_\pi^6}{1260\,M^6} 
+ \frac{193\,m_\pi^8}{2520\,M^8}\right)
 \nonumber \\ & &  
-\frac{m_\pi^2}{20\,M^4}\left(1 + \frac{9\, m_\pi^2}{7\, M^2}
+ \frac{76\, m_\pi^4}{63\, M^4}\right)(s+t')
 \nonumber \\ & & 
	+\frac{1}{60\,M^4}\left(1 + \frac{9\, m_\pi^2}{7\, M^2}
        + \frac{94\, m_\pi^4}{63\, M^4}\right)(s^2+t'^2)
	+\frac{1}{60\,M^4} \left(1 + \frac{2\, m_\pi^2}{M^2}
	+ \frac{341\, m_\pi^4}{126\, M^4}\right) s\,t'
 \nonumber \\ & &
- \frac{1}{252\, M^6}\left(1+\frac{29\, m_\pi^2}{10\, M^2}\right)
	(s^2 t' + s t'^2 )
        - \frac{m_\pi^2}{315\, M^8}(s^3+t'^3)
 \nonumber \\ & &
	+\frac{1}{1890\, M^8}\left( s^4+ 2 s^3 t' + 3 s^2 t'^2
	+2 s\, t'^3 +t'^4\right) + {\cal O}(p^{10}) \; .
\end{eqnarray}

At CEBAF \cite{Miskimen+al94}, one takes data near 
$t\approx -m_\pi^2$, for which our ${\cal O}(p^{8})$ expansion
yields
\begin{eqnarray}\label{eqmisk}
\widetilde{F}^{3\pi}_\gamma(s,t')&=&\left( 
1 + \frac{m_\pi^2}{6\,M^2} + \frac{m_\pi^4}{20\,M^4} + 
  \frac{13\,m_\pi^6}{1260\,M^6} + \frac{\,m_\pi^8}{360\,M^8} \right)
	-\frac{m_\pi^2}{60\,M^4} \left(1 + \frac{m_\pi^2}{3\,M^2}
	+ \frac{8\,m_\pi^4}{63\,M^4}\right)s
 \nonumber \\ & &
	-\frac{m_\pi^2}{60\,M^4} \left(1 +  \frac{3\,m_\pi^2}{7\,M^2}
        + \frac{10\,m_\pi^4}{63\,M^4}\right)t'
	+\frac{1}{60\,M^4} \left(1 + \frac{3\,m_\pi^2}{7\,M^2}
	+ \frac{13\,m_\pi^4}{63\,M^4}\right)s^2
\nonumber \\ & &
	+\frac{1}{60\,M^4} \left(1 + \frac{3\,m_\pi^2}{7\,M^2}
        + \frac{4\,m_\pi^4}{21\,M^4}\right)t'^2
	+\frac{1}{60\,M^4} \left(1 + \frac{2\,m_\pi^2}{3\,M^2} +
  \frac{41\,m_\pi^4}{126\,M^4}\right)s\,t'
 \nonumber \\ & &
- \frac{1}{252\, M^6}\left(1+\frac{29\, m_\pi^2}{30\, M^2}\right)
        (s^2 t' + s\, t'^2 )
        -\frac{m_\pi^2}{945\,M^8} (s^3+t'^3)
 \nonumber \\ & &
	+\frac{1}{1890\, M^8}\left( s^4 +2 s^3 t' + 3s^2 t'^2
	+2 s\, t'^3 + t'^4\right) + {\cal O}(p^{10}) \, ,
\end{eqnarray}
where the $s\leftrightarrow t'$ symmetry is lost due to
the virtuality $t=-p_1^2=-p_2^2= -m_\pi^2$ of the third pion.

Our momentum expansions (\ref{eqant}) and (\ref{eqmisk}) show 
clearly that the main contribution to the term linear in $s$ and 
$t'$ (dominating the $s,t'$-dependence close to the soft limit), 
comes from ${\mathcal{O}}(p^4)$ and not ${\mathcal{O}}(p^2)$. 
This happens since the ${\mathcal{O}}(p^2)$-terms,
due to the constraint (\ref{oscond}), contribute
only to the part independent of $s$ and $t'$. 

Note that in the on-shell case (\ref{eqant}), the finite pion mass 
$m_\pi$ causes a larger upward shift than in the off-shell 
case (\ref{eqmisk}).
For chiral pions ($m_\pi=0$) and real photons the condition 
(\ref{oscond}) becomes $s+t'+u= t$. For this case, 
but for general $t$, the amplitude (\ref{sumF}) becomes 
\begin{eqnarray}\label{eqmisk2}
\tilde{F}^{3\pi}_\gamma(s,t')&=& \left(1+\frac{t}{6\,M^2}
+\frac{t^2}{30\,M^4} +
  \frac{t^3}{140\,M^6} + \frac{t^4}{630\,M^8} \right)
-\frac{t}{60\,M^4} \left(1 + \frac{3\,t}{7\,M^2}
        + \frac{t^2}{7\,M^4}\right)(s+t')
 \nonumber \\ & &
+\frac{1}{60\,M^4} \left(1 + \frac{3\,t}{7\,M^2}
        + \frac{11\,t^2}{63\,M^4}\right)(s^2+t'^2)
+\frac{1}{60\,M^4} \left(1 + \frac{2\,t}{3\,M^2} +
  \frac{13\,t^2}{42\,M^4}\right)s\,t'
 \nonumber \\ & &
- \frac{1}{252\, M^6}\left(1+\frac{29\, t}{30\, M^2}\right)
        (s^2 t' + s\, t'^2 )
-\frac{t}{945\,M^8}\, (s^3+t'^3)
 \nonumber \\ & &
+\frac{1}{1890\, M^8}\left( s^4 +2 s^3 t' + 3s^2 t'^2
        +2 s\, t'^3 + t'^4\right) + {\cal O}(p^{10}) \, ,
\end{eqnarray}
showing that the $s\leftrightarrow t'$ symmetry is restored
in the chiral limit. 
The massless pion amplitude (\ref{eqmisk2}) is smaller than 
the one with $m_\pi = 138.5$ MeV by, typically, 4\% when 
$M=m_\rho/2 = 385$ MeV, 
by some (depending on $s$) 6\% when $M=330$ MeV, by more than
10\% when $M=250$ MeV, {\it etc}.
It is interesting that for small $s$ and $t'$, 
the chiral-limit $F^{3\pi}_\gamma$ can fall slightly below
its soft point value (\ref{g3piAnomAmp}) when $t < 0$.

In the CEBAF experiment \cite{Miskimen+al94}, $s$ will vary 
more than $t'$. 
For the $t'$-range relevant at CEBAF, the $t'$-dependence of
$F^{3\pi}_\gamma$ anyway turns out to be rather weak. For example, 
suppose one plots (not done here to avoid overcrowding our figures) 
the $t=-m_\pi^2$ form factor (\ref{eqmisk}) as a function of $s$ 
for several values of $t'$ varying from $t'=-m_\pi^2$ to 
$t'=-8m_\pi^2$. (Take $M=330$ MeV for definiteness.) 
One would thus get, across the whole $s$-range relevant at CEBAF, 
a narrow strip of tightly spaced curves, where the curve
depicting $\widetilde{F}^{3\pi}_\gamma(s)$ for $t'=-8m_\pi^2$, 
would differ by just 2\% to 3\% from the curve with $t'=-m_\pi^2$.
In Figs. 2 and 3, we therefore show the dependence of 
$F^{3\pi}_\gamma$ (for various cases) on the variable $s$,
with $t'$ fixed. 
On the other hand, since $t$ changes sign when we put also the 
third pion off-shell as at CEBAF \cite{Miskimen+al94}, the amplitude 
is more sensitive to this change. 

%\section{Discussion, Comparisons with Other Models, and Conclusions}

In Fig. 2 we give our results at $t=-m_\pi^2$ (the $t$-value
most relevant at CEBAF \cite{Miskimen+al94})
for the constituent masses $M=385$ MeV, 330 MeV, 300 MeV and 
$M=250$ MeV. In fact, we show a pair of curves for each of these 
masses: one of the curves is obtained by exact numerical evaluation 
of the box amplitude, and the other corresponds to our series 
expansion to the order ${\cal O}(p^8)$. 
At the lowest depicted $s$, the form factor
obtained by the accurate numerical calculation
is (for each $M$) slightly below the corresponding
series expansion approximating it, but exceeds it
eventually as $s$ grows.
The convergence of the expansion is very satisfactory on the whole,
since for 
$s \raise 0.2em\hbox{$<$}\kern -0.8em \lower 0.25em\hbox{$\sim$} 
11 m_\pi^2$, 
the agreement between the two ways of calculation is very good for 
all these values of $M$. For $s > 11 m_\pi^2$, the exception are 
only the cases with unrealistically small $M$, such as $M=250$ MeV. 
For $M=250$ MeV, we plot the curves up to 
$s=s_{\rm tr}\approx 13.03 m_\pi^2$ only.
Namely, when $s$ reaches $s_{\rm tr}=(2M)^2$, {\it i.e.},
the threshold for production of an on-shell quark-antiquark pair,
the QL approach becomes inadequate, because the amplitude starts
being dominated by this threshold which is not physical but an 
artifact of the model (\ref{1Leff}).
[Concerning the accuracy of the computations close to a threshold:
the difficulty of numerical integration starts increasing gradually
as one gets closer than $m_\pi^2$ to the threshold, while the 
accuracy of the ${\mathcal{O}}(p^8)$ expansion starts failing 
before that.]

Nevertheless, for the values of $s < 16 m_\pi^2$
accessible at CEBAF \cite{Miskimen+al94}, such a 
threshold cannot be reached unless $M < 2 m_\pi = 277$ MeV.
Such values are, however, too low to serve as the constituent 
quark masses, which cannot be much lighter than 
$M \approx M_{nucleon}/3 \approx 313$ MeV.

In Fig. 3 we compare our (numerically obtained) on-shell 
predictions for various $M$ with the only existing experimental 
point so far \cite{Antipov+al87}, and with the 
predictions of chiral perturbation theory ($\chi$PT) 
\cite{Bijnens90} (in Holstein's \cite{Holstein96} renormalization 
convention -- {\it i.e.}, we take his \cite{Holstein96} Eq. (10)
for the $\chi$PT form factor), of vector meson dominance (VMD)
\cite{Rudaz84} [Holstein's \cite{Holstein96} Eq. (9)], and of 
VMD with the final pion rescattering included \cite{Holstein96}. 

We conclude that our results agree rather well with VMD for 
constituent quark masses $300\;\mathrm{MeV} < M < 330\;\mathrm{MeV}$
when $s > 8 m_\pi^2$. Going down in $s$, already at $s\sim 8 m_\pi^2$
we agree rather well, but for somewhat higher $M$ (330 MeV 
$\raise 0.2em\hbox{$<$}\kern -0.8em \lower 0.25em\hbox{$\sim$} M 
\raise 0.2em\hbox{$<$}\kern -0.8em \lower 0.25em\hbox{$\sim$} 
m_\rho/2$), with both VMD and $\chi$PT.
For all $s$-values shown, our predictions get somewhat closer to 
those of $\chi$PT when the ratio $m_\pi^2/M^2$ gets smaller, 
{\it i.e.}, for the largest considered constituent quark mass,  
$M = m_\rho/2 = 385$ MeV. 
Since $\chi$PT results in the weakest momentum dependence, its 
predictions for the largest values of $s$ accessible at CEBAF
are significantly different from both VMD and our QL ($\chi$QM, 
$\sigma$-model) approach. The CEBAF experiment 
should thus be able to distinguish between these various 
physical mechanisms in this range of momenta. On the other
hand, all these approaches agree reasonably well for the lowest 
of $s$-values 
accessible at CEBAF. In particular, our approach agrees with  
VMD and $\chi$PT that the existing data point \cite{Antipov+al87}
is probably an overestimate, as we can fit it well ({\it e.g.},
with the $M=250$ MeV curve in Fig. 3) only for unrealistically 
low values of $M$. 

\vskip 8mm

%\section*{Acknowledgments}
The authors thank R. Alkofer and D. Kekez for many helpful 
discussions, and K. Kumeri\v cki for checking the manuscript. 
The support of the Croatian Ministry of Science and Technology 
contract 1--19--222 is also acknowledged.

%%%%%%%%%%%%%%%%%%%%%%% end of freebib.tex %%%%%%%%%%%%%%%%%%
\newpage

\section*{Figure captions}

\begin{itemize}

\item[{\bf Fig.~1:}] One of the six box diagrams for the process 
$\gamma \to \pi^+ \pi^0 \pi^-$. The position of the $u$ and $d$ 
quark flavors on the internal lines, as well as $Q_u$ or $Q_d$ 
quark charges in the quark-photon vertex, varies from graph to 
graph, depending on the position of the quark-pion vertices. 

\item[{\bf Fig.~2:}]
Our numerically obtained $\gamma\to 3\pi$ form factors are compared 
with the corresponding ${\mathcal{O}}(p^8)$-expansions (\ref{eqmisk})
for various values of $M$ and $m_\pi = 138.5$ MeV. The corresponding 
pair of curves for $M=385\;\mathrm{MeV}$ is denoted by the solid 
lines, by the short-dashed lines for $M=330\,\mathrm{MeV}$, by the 
long-dashed lines for $M=300\, \mathrm{MeV}$, and by the very 
short-dashed lines (only up to $s_{\rm tr}=13.03\, m_\pi^2$,
denoted by $\bf x$'s) for $M=250\, \mathrm{MeV}$.
The curves resulting from numerical integration start slightly 
below their respective power series at lowest values of the 
$s$-variable, but then exceed the latter for sufficiently high 
$s$. All curves pertain to the off-shell case $t=-m_\pi^2$ 
(dominant at CEBAF \cite{Miskimen+al94}, but not in the Serpukhov 
experiment \cite{Antipov+al87} which provided the displayed data 
point, where $t=m_\pi^2$). The remaining variable $t'$ is also set 
to $t' = -m_\pi^2$.

\item[{\bf Fig.~3:}] 
The ${\widetilde F}_\gamma^{3\pi}(s,t'=-m_\pi^2)$ 
predicted by VMD \cite{Rudaz84,Holstein96} (solid curve), 
VMD with final pion interactions \cite{Holstein96} (dotted curve), 
and $\chi$PT \cite{Bijnens90,Holstein96} (dash-dotted curve), 
are compared with our ${\widetilde F}_\gamma^{3\pi}(s,t'=-m_\pi^2)$ 
obtained by numerical integration for $M=330$ MeV (short-dashed 
curve), $M=300$ MeV (long-dashed curve), and $M=250$ MeV (the 
topmost, very short-dashed curve). Same as the displayed data 
point \cite{Antipov+al87}, all curves pertain to all three pions 
on shell, so that $t=m_\pi^2=(138.5$ MeV)$^2$. The remaining free 
variable, $t'$, is set to $t' = -m_\pi^2$ for all curves for 
definiteness. 

\end{itemize}

%%%%%%%%%%%%%%%%%%%%%%%%%%%%%% FIGURES %%%%%%%%%%%%%%%%%%%%%%%%%%%
\newpage

\epsfig{file=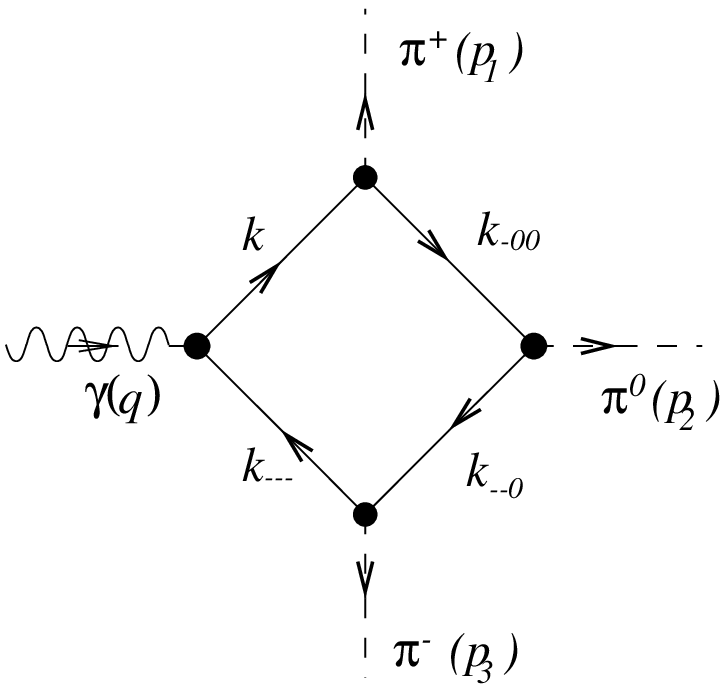,width=15cm}

\newpage
\epsfig{file=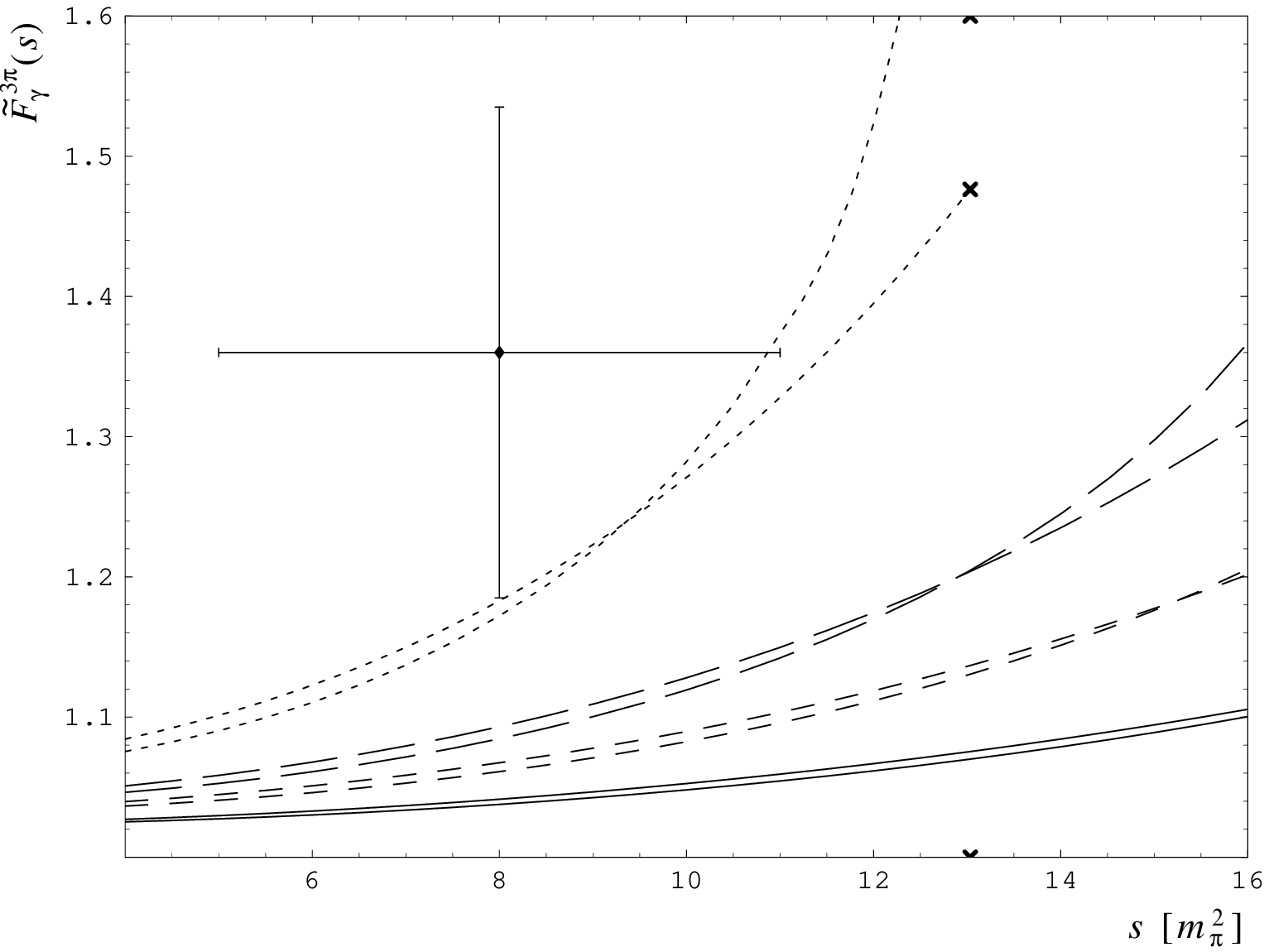,width=15cm}

\newpage
\epsfig{file=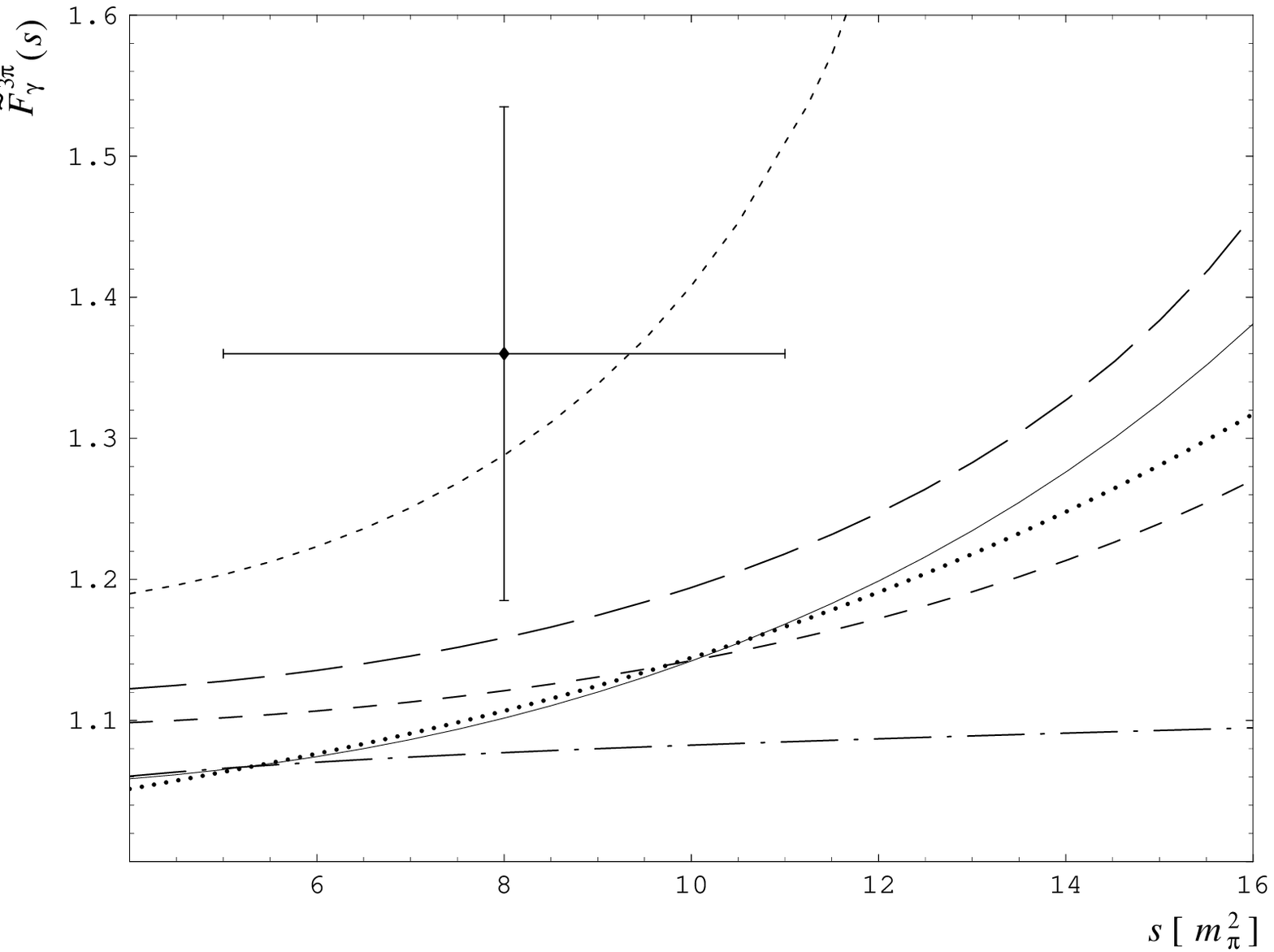,width=15cm}

%%%%%%%%%%%%%%%%%%% end of freepic.tex %%%%%%%%%%%%%%%%%%%%%%
\end{document}